\documentclass[twocolumn]{aa}
\usepackage{booktabs}   
\usepackage{arydshln} 
\usepackage{dblfloatfix}
\usepackage{amsmath}
\usepackage{txfonts}
\usepackage{booktabs}
\usepackage{float}
\usepackage{amsmath}
\usepackage[utf8]{inputenc}
\usepackage{graphicx}
\usepackage{float}
\usepackage{hyperref}
\usepackage{csvsimple}
\usepackage{adjustbox}
\usepackage{natbib}
\begin{document}

\title{3D kinematics of SMC star clusters: residual velocities \\disentangle
kinematically perturbed clusters}

\author{Denis M.F. Illesca\inst{1,2}\thanks{\email{denisillesca1113@gmail.com}}
\and  Andr\'es E. Piatti\inst{1,2} \and  Mat\'{\i}as Chiarpotti\inst{1,2} \and Roberto Butr\'on\inst{1}}

\institute{Instituto Interdisciplinario de Ciencias B\'asicas (ICB), CONICET-UNCuyo, Padre J. Contreras 1300, M5502JMA, Mendoza, Argentina;
\and Consejo Nacional de Investigaciones Cient\'{\i}ficas y T\'ecnicas (CONICET), Godoy Cruz 2290, C1425FQB,  Buenos Aires, Argentina\\
}

\date{Received / Accepted}

\abstract{Understanding the kinematic behaviour of the Small Magellanic Cloud (SMC) remains a challenge addressed by many authors using diverse approaches. Over time, increasing observational evidence has accumulated for tidal perturbations induced by the Large Magellanic Cloud (LMC) on the SMC, especially in its outer regions. 
In this study, we adopt star clusters as kinematic tracers of the SMC. We analyse 36 clusters distributed across the galaxy's structural regions (Northern Bridge, Southern Bridge, Wing/Bridge, West Halo, Main Body and Counter-Bridge). From each cluster's proper motions, radial velocity and heliocentric distance we estimate Cartesian velocities \((V_x,\,V_y,\,V_z)\) in the SMC reference frame. We also compute the same velocity components under the assumption that the SMC behaves as a rotating disc. 
We then define the residual velocity \(\Delta V\) for each cluster as the difference between the two velocities derived. Additionally, we perform a kinematic anisotropy analysis to characterise the distribution of kinetic energy across the SMC.

We find that increasing values of \(\Delta V\) correlate with increasing cluster distance from the SMC center, and that \(\Delta V \approx 60\ \mathrm{km\,s^{-1}}\) it appears to be a lower limit that separates, in kinematic terms, the areas of tidal origin from those with the best behavior.} 
 
\keywords{(Galaxies:) Magellanic Clouds – Galaxies: kinematics and dynamics - Galaxies: star clusters: general}

\titlerunning{SMC kinematic}

\authorrunning{D.M.F. Illesca et al.}

\maketitle

\markboth{D.M.F. Illesca et al.: }{SMC star clusters}

\section{Introduction}

The kinematic behavior of the Small Magellanic Cloud (SMC) has been investigated in recent 
years by several authors  
\citep[e.g.,][]{kallivayalil2013third, zivick2018proper,de2020revealing, niederhofer2021vmc, dhanush2025unraveling}.Understanding the internal kinematics of the
SMC is essential for reconstructing its interaction with  the Large Magellanic Cloud 
and the Milky Way.

To trace the kinematic signatures of the SMC, different galactic constituents have been used,
namely:
HI gas, young stars, red giant stars, and massive stars, among others. The gas in
the SMC exhibits considerable internal rotation \citep{stanimirovic2004new}, while young stars 
show an orderly motion towards the Magellanic Bridge, with proper motions
greater than that of the SMC main body \citep{oey2018resolved}. Indeed,
\citet{nakano2025evidence} investigated 
the motions of massive stars ($>$\,8\,M\(_\odot\)) with ages  <\,50\,Myr, and found
trajectories oriented towards the LMC and away from the SMC main body. In contrast, the
 oldest stellar population apparently shows little rotation 
\citep{harris2006spectroscopic,zivick2021deciphering}, which make the whole
SMC kinematics - to some extent - a still living conundrum.  We note, however, that some of 
these results are based solely on radial velocity or proper motion measurements.

The SMC is under tidal effects due to its interaction with the LMC \citep{mackey2018substructures,zivick2018proper,de2020revealing,niederhofer2021vmc,omkumar2021gaia}. 
The magnitude and strength of tidal forces on the morphology and internal kinematics of the SMC were
estimated from dynamic simulations by \citet{besla2012role}. They concluded that the 
Magellanic Clouds are in their first fall towards the Milky Way. 
In this context, \citet{piatti2021kinematics} used star clusters as tracers of the internal 
kinematics of the SMC and constructed a 3D image of the clusters' motions from {\it Gaia} data
\citep{gaiaetal2016} and radial velocities obtained from the literature. The cluster motions 
derived by \citet{piatti2021kinematics} show some notable dispersion around the resulting
rotating disk. This finding reveals that the kinematics of the SMC clusters is complex and cannot 
be fully captured by a representation of a rotating disk alone.

In this work, we analyze 36 SMC star clusters with the aim of obtaining a 
comprehensive representation of the SMC’s internal kinematics, based on heliocentric distances 
obtained by \citet{illesca2025astrophysical}, proper motions retrieved from {\it Gaia} Data Release 3 \citep{gaiaetal2016,luri2021gaia}, and radial velocities available in the literature.
Incorporating individual cluster heliocentric distances, rather than adopting a single SMC mean distance 
for all the clusters, makes the derived kinematic behaviors more robust. From this data set, we 
construct a three-dimensional velocity map of the SMC, following the formalism of \citet{van2002new}.
We then analyze the residual velocities and explore the resulting kinematic signatures 
across known tidally perturbed SMC structures.
In Section 2, we describe the data collected and employed in the present analysis. In Section 3, 
we describe  the results obtained, while in Section 4 we discuss the residual velocities of star clusters 
as indicators of kinematic perturbations caused by tidal forces.
Section 5 summarizes the main conclusions of this work.

\section{Data collection and processing}

\citet{illesca2025astrophysical} studied 40 SMC star clusters, mainly distributed across the
outer SMC regions with the aim of investigating the connection between
their ages, heliocentric distances and metallicities. We used their cluster collection as a 
starting point to build a sample of SMC star clusters with the three mentioned fundamental
parameters, in addition to radial velocities (RVs) and proper motions. Unfortunately,
as far as we are aware, 12 star clusters do not have RVs available in the literature
(B88, B139, BS116, HW64, HW67, HW73, HW77, IC1655, L2, L3, L73, and L95). In contrast,
we added other 8 star clusters with the required information (L1, L8, L12, L68, L113, NGC~339, 
NGC~361, and NGC~419), with their fundamental parameters taken from \citet{piatti2023depth}. 
For the final sample of 36 star clusters (accurate individual cluster heliocentric distance 
was required), we extracted the clusters' right ascension (RA), declination (Dec.), and
radii from \citet{bica2020vizier}. As for the astrometric information, we
retrieved from {\it Gaia} DR3 proper motions in 
right ascension (pmra), proper motions in declination (pmdec), parallaxes \( \varpi \), excess 
noise \textbf{(epsi)}, significance of excess noise \textbf{(sepsi)}, and $G$, $BP$, and $RP$
magnitudes for every star located within three times the respective cluster's radius.
We applied a filter to the proper motion errors to retain those stars with $\sigma$
\( \leq 0.1~\mathrm{mas}~\mathrm{yr}^{-1} \), following the procedure described in \citet{piatti2019two}.
We favored the selection of extragalactic stars by applying the condition 
\( |\varpi| / \sigma(\varpi) < 3 \). Furthermore, in order to improve our data quality, we limited \textbf{sepsi}\( < 
\)2, \textbf{epsi}\( < \)1, \textbf{RUWE} $\le$ 1.4, and $G$ $\le$ 18 mag, respectively
\citep[see, e.g.][]{ripepi2019reclassification}

\begin{figure}
\includegraphics[width=\columnwidth]{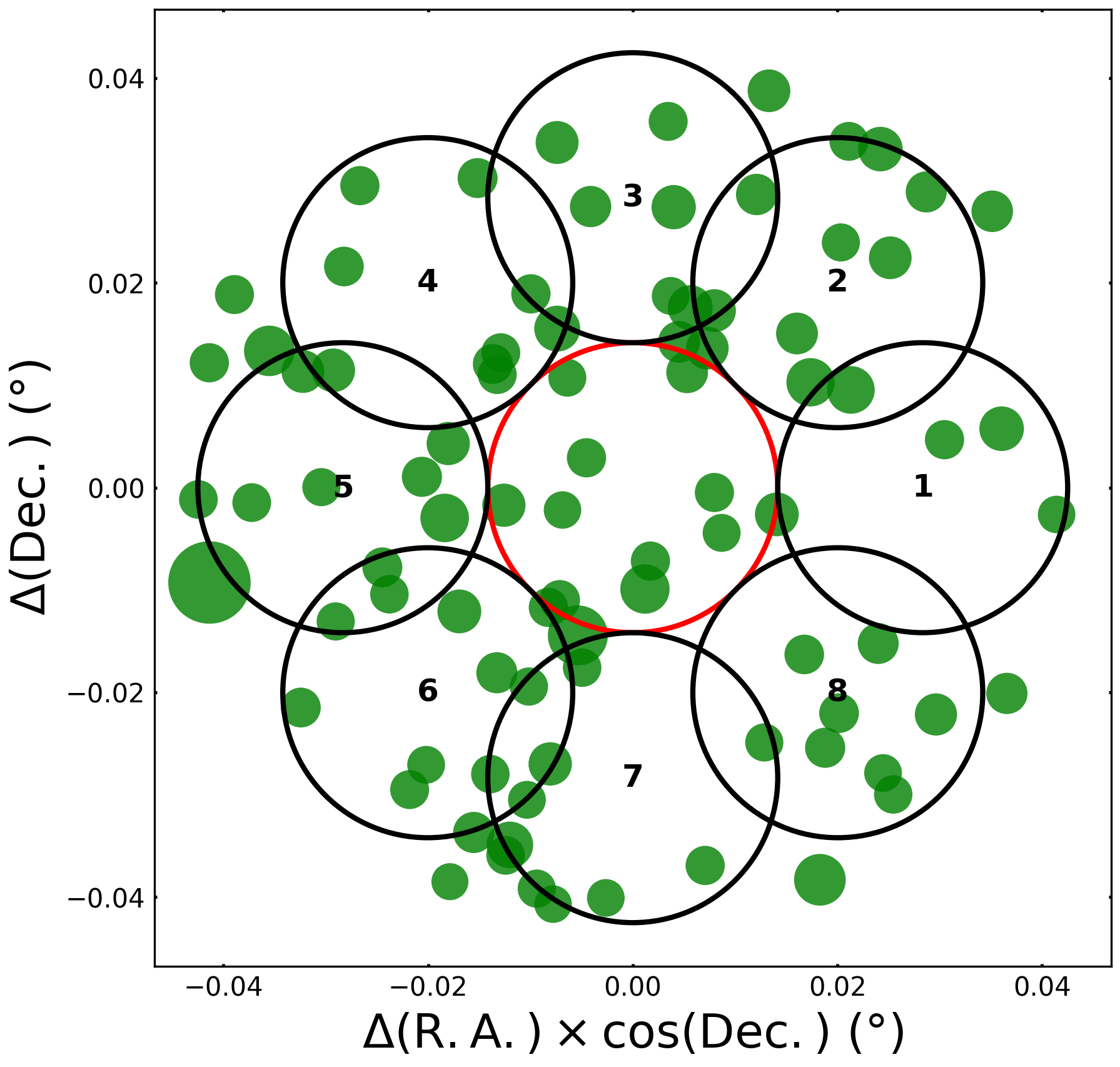}
\caption{Stars selected from {\it Gaia} DR3 distributed in the field of NGC~416. The red circle 
corresponds to the cluster field, while the black circles correspond to 8 different comparison fields placed 
adjacent to the cluster. The size of the symbols is proportional to the brightness of the stars
in the $G$ filter.}
\label{fig:ngc416_limp}
\end{figure}

\begin{figure} 
\includegraphics[width=\columnwidth]{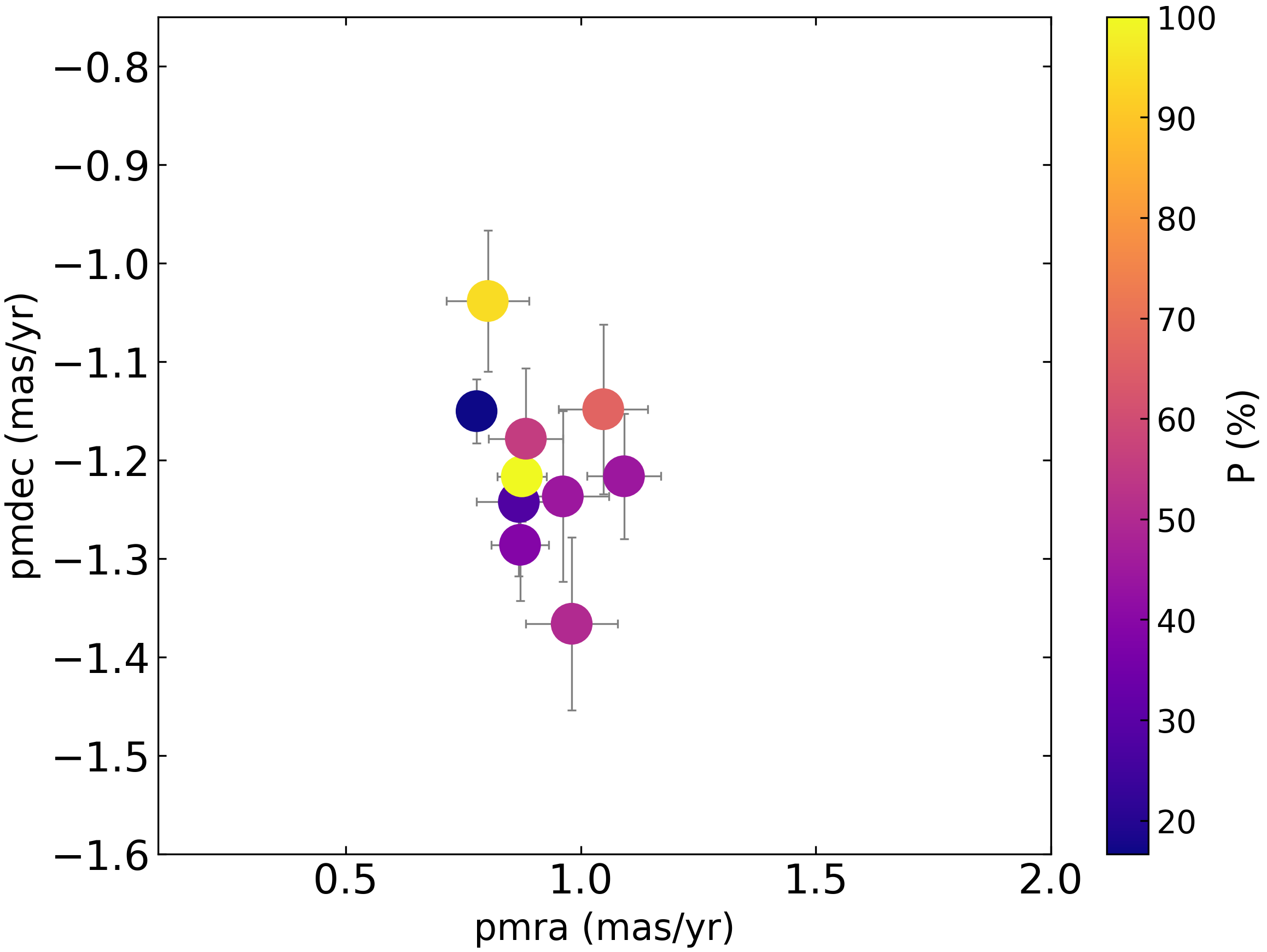}
\caption{VPD for selected {\it Gaia} DR3 stars distributed in the field of NGC~416. Color symbols
vary according to the assigned membership probability.}
\label{fig:ngc416_vpd}
\end {figure}

We then used the procedure devised by \citet{piatti2012washington}, originally designed to clean 
star cluster color-magnitude diagrams from field star contamination, to statistical remove SMC 
field stars from the
vector point diagrams (VPDs) of the star clusters. The statistical cleaning method 
makes use of comparison field regions surrounding each cluster. Figure~\ref{fig:ngc416_limp}
illustrates the locus of the cluster circle with respect to 8 different circular 
comparison fields of the same area as the cluster's circle. The method superimposes the cluster and
one comparison field VPDs, and for each star in the latter it subtracts the closest one
in the cluster VPD.  The proper motion errors of the stars were also taken into account 
when searching for a star to subtract from the cluster's VPD. To do this, we allowed the proper 
motions of the stars in the cluster's VPD to vary within a range of \( \pm 1 \sigma \). 
We repeated the procedure described above for one thousand comparison fields
placed around the cluster's circle at randomly chosen position angles. Finally, we assigned 
to the stars in the cluster's VPD probabilities of being cluster members as \(P\,(\%) = 100 \times S / 1000\), 
where S represents the number of times the star has not been subtracted after a thousand different 
runs. An example of the results obtained is illustrated in  
Figure~\ref{fig:ngc416_vpd}. Stars with different $P$ values were plotted with different 
colors. In the subsequent analysis, we retained only stars with \( P > 50\% \).

For the number $N$ of stars that satisfy the above restriction in each cluster, we applied the concept of effective sample size introduced by \citet{kish1987weighting}. We defined an effective 
$N$ ($N^{\mathrm{eff}}$) as a representative and comparable measure of the star-by-star kinematics. We computed
$N^{\mathrm{eff}}$ using the expression:\\

\begin{equation}
N^{\mathrm{eff}} =
\frac{\widetilde{N}}{\max(\widetilde{N})},
\label{eq:neff_formulas}
\end{equation}

\noindent where:

\begin{equation}
\widetilde{N} =
\bar{P} \,\sqrt{N} \,\frac{\left(\sum_{k=1}^{N} p_{k}\right)^{2}}{\sum_{k=1}^{N} p_{k}^{2}}
\qquad\qquad.
\end{equation}

In Eq. (2), $\widetilde{N}$ combines the individual star membership probabilities $p_{k}$, $k=1,...,N$, 
with an average value $\bar{P}$. 
The first factor $\bar{P}$ in Eq. (2) penalizes star clusters with averaged low membership probabilities,
while the second factor $\sqrt{N}$ penalizes clusters with smaller numbers of stars; the
ratio corresponds to the 
\citet{kish1987weighting}'s effective sample size.  $N^{\mathrm{eff}}$ is the normalized 
version of $\widetilde{N}$  and we used it in the subsequent analysis 
as a relative quality weight in our star cluster kinematic results.

We applied a maximum likelihood statistical method \citep{meylan1993observational,walker2006internal} 
to estimate the mean proper motions and dispersion of the studied clusters. In practice, 
we optimized the probability $\mathcal{L}$ such that a given set of stars with proper motions 
(\( pm_{i} \)) and errors \( \sigma_{i} \) is extracted from a population with mean proper motion 
\( \langle pm \rangle \) and dispersion W, as follows: \[
\mathcal{L} = \prod_{i=1}^{N} \left( 2 \pi \left( \sigma_i^2 + W^2 \right) \right)^{-\frac{1}{2}} 
\exp\left( 
    - \frac{ \left( pm_i - \langle pm \rangle \right)^2 }{ 2 \left( \sigma_i^2 + W^2 \right) }
\right),
\]
where the mean and dispersion errors were calculated from the respective covariance matrices.  
The resulting mean cluster proper motions are shown in Table~\ref{tab1}, alongside the number
of stars ($N$) used to compute them.

\section{Star cluster kinematic properties}
We choose star clusters as kinematic tracers because they provide
with a robust methodology that distinguishes it from other approaches. 
Unlike the selection of field star populations \citet{dhanush2025unraveling}, clusters are discrete, gravitationally bound objects. This allows estimating their ages, distances and
velocities with a greater accuracy than for field stars.
Moreover, our star cluster sample includes individual heliocentric distances, which 
constitute a valuable feature compared to kinematic models based on field stars that employ mean distances, thereby underestimating the role of distances.

Stellar clusters, as kinematic tracers, provide an insightful view of the SMC
kinematic, without the biases that different tracers might introduce because of lack
of distance estimates. Although \citet{dhanush2025unraveling} perform a differential analysis by populations to account for changes in geometry and, consequently, in the kinematic model, 
we here exploit the SMC kinematic model obtained by \citet{piatti2021kinematics} which is 
based on star clusters. He found that the SMC rotation disk is characterized by the right ascension and declination of its center (RA = $13.30^\circ \pm 10$, Dec = $-72.85^\circ \pm 10$), its distance to the center ($59 \pm 1.5\ \mathrm{kpc}$), radial velocity ($150 \pm 2\ \mathrm{km\ s^{-1}}$), central proper motion in RA ($\mathrm{pmra_{center}} = 0.75 \pm 10\ \mathrm{mas\ yr^{-1}}$), central proper motion in Dec ($\mathrm{pmdec_{center}} = -1.26 \pm 0.05\ \mathrm{mas\ yr^{-1}}$), disk inclination ($70^\circ \pm 10$), position angle of the line of nodes ($200 \pm 30$), and rotation velocity ($25 \pm 5.0\ \mathrm{km\ s^{-1}}$), respectively.

We firstly subtracted the mean proper motion and radial velocity of the SMC center of mass \citep{piatti2021kinematics} from the resulting clusters' mean proper motions and radial velocities,
and calculated the residual linear velocities V$_{\rm RV}$, V$_{\rm RA}$ and V$_{\rm Dec}$, 
the latter in units of [$\mathrm{km~s^{-1}}$] through the expression 
4.7403885 $\times$ $D$~[$\mathrm{mas~yr^{-1}}$], where $D$ is the cluster heliocentric distance.

To convert the vector (V$_{\rm RV}$, V$_{\rm RA}$, V$_{\rm Dec}$) into one with components $Vx$ 
and $Vy$ in the plane of the SMC 
and $Vz$ perpendicular to it, we used  the reference system defined by \citet{van2002new}, and
followed the procedure described in \citet{piatti2019two}. This comprised inverting the 
matrix \textbf{A = B $\boldsymbol{\times}$ C}, where \textbf{B} is the matrix: \[
\begin {pmatrix}
1 & 0 & 0 \\
0 & b_1 & b_2 \\
0 & b_3 & b_4
\end{pmatrix}
\] with \( b_{1} \),  \( b_{2} \),  \( b_{3} \), and  \( b_{4} \) being the coefficients of 
the transformation Eq. (9), and \textbf{C} is the matrix defined in Eq. (5) of 
\citet{van2002new}, respectively, so that :

\begin {equation}
\begin{pmatrix} V_x \\ V_y \\ V_z \end{pmatrix} = \mathbf{A}^{-1} \begin{pmatrix} \mathrm{V_{\rm RV}} 
\\ \mathrm{V_{\rm RA}} \\ \mathrm{V_{\rm Dec}} \end{pmatrix}
\end{equation} 

\noindent The errors $\sigma(V_x)$, $\sigma(V_y)$ and $\sigma(V_z)$ were estimated by
performing Monte Carlos experiments using the uncertainties in  V$_{\rm RV}$, V$_{\rm RA}$ 
and V$_{\rm Dec}$. From Eq. (3) we calculated $\mathrm{V_{rot}} = (V_{x}^2 + V_{y}^2)^{1/2}$
and $\mathrm{V_{rot.3D}} = (V_{x}^2 + V_{y}^2 + V_z^2)^{1/2}$, and the resulting values
are listed in Table~\ref{tab2}.

On the other hand, we computed the velocity components ($V_{x'}$, $V_{y'}$, $V_{z'}$) with
respect to the SMC center that the star clusters would have, if they rotated at their present 
positions in the SMC disk according to the rotation disk fitted by \citet{piatti2021kinematics}.
The difference between ($V_{x}$, $V_{y}$, $V_{z}$) and ($V_{x'}$, $V_{y'}$, $V_{z'}$) is the
so-called residual velocity vector ($\Delta V_x$, $\Delta V_y$, $\Delta V_z$), 
where  $\Delta V_x$ = $V_x - V_{x'}$, $\Delta V_y$ = $V_y - V_{y'}$, and $\Delta V_z$ = 
$V_z - V_{z'}$, respectively. The resulting values are listed in Table~\ref{tab3}.
The module of the residual velocity vector 
($\Delta V = (\Delta V_x^2 + \Delta V_y^2 + \Delta V_z^2)^{1/2}$)
was introduced by \citet{piatti2021residual} as a measure of the kinematic perturbation
experienced by a star cluster, i.e, how much the cluster's motion departs from an ordered
rotation.

\begin{figure*}  
  \centering
  \includegraphics[width=\textwidth]{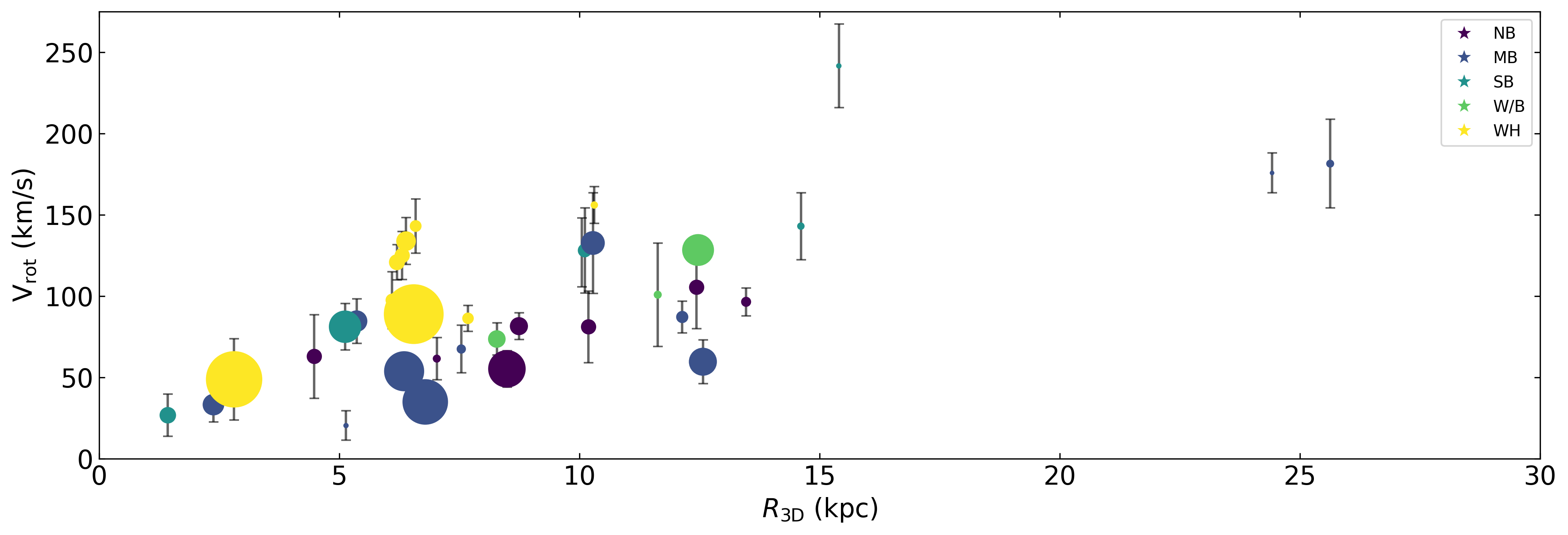}
  \caption{Rotational velocity values for the studied star clusters as a
function of their distances to the SMC's center. The colors represent different structures of the SMC
 \citep{dias2016smc} as indicated in the top-right panel (NB=Northern Bridge, 
MB= Main Body, SB=Southern Bridge, W/B=Wing/Bridge, WH=West Halo). Symbol size is proportional to $N^{\mathrm{eff}}$.}
  \label{fig:velrot}
\end{figure*}

Finally, following \citet{van2001magellanic}, we computed the Cartesian coordinates
($x,y,z$) of the star clusters with respect to the SMC's center:

\begin{equation}
\begin{aligned}
x &= D \sin \rho \cos (\phi - \theta), \\
y &= D \left[ \sin \rho \cos i \sin (\phi - \theta) + \cos \rho \sin i \right] - D_0 \sin i, \\
z &= D \left[ \sin \rho \sin i \sin (\phi - \theta) - \cos \rho \cos i \right] + D_0 \cos i.
\end{aligned}
\label{eq:coords_transform}
\end{equation}

\noindent where $D$, $\rho$ and $\phi$ are the cluster heliocentric distances, the cluster projected distances from the SMC's center and their position angles, respectively, the latter computed from the
cluster celestial coordinates (RA, Dec.). $D_O$ represents the mean heliocentric distance
of the SMC's center \citep[62.44 kpc, ][]{graczyk2020distance}, while $\theta$ and $i$ are the
position of the line of nodes  and the inclination of the SMC disk derived by \citet{piatti2021kinematics}. From Eq. (4), we computed the projected distance on the SMC plane
$R_{\mathrm{plane}} = \sqrt{x^2 + y^2}$, and the space distance $R_{\mathrm{3D}} = 
\sqrt{x^2 + y^2 + z^2}$, and listed the resulting values in Table~\ref{tab3}. 
At first glance, we found that most of the selected star clusters are
distributed within $R_{\mathrm{3D}} \sim$ 14 kpc, some few ones reaching $R_{\mathrm{3D}}\sim$ 
25 kpc (see Figure~\ref{fig:velrot}). 

\section{Analysis and discussion}

\citet{besla2012role} showed that the irregular morphology and internal kinematics of 
the Magellanic System can more robustly explained by considering gravitational interactions 
between the LMC and the SMC. This outcome leads to question about the kinematic 
signatures witnessing the tidally disturbed structures of the SMC.
We here addressed this issue by using star clusters as kinematic tracers, and their
residual velocities as a measure of the perturbed kinematic signatures. In this context, 
star clusters located in tidally perturbed SMC regions are expected to have larger residual 
velocities. For instance, \citet[][see his Figure~3]{piatti2021kinematics} found that 
star clusters pertaining to outer SMC regions (some of them with a known tidal origin)
have $\Delta V >$ 50~$\mathrm{km\,s^{-1}}$. We built a similar figure (see Figure~\ref{fig:deltav})
using our sample of 36 star clusters. As can be seen, star clusters located outside the
SMC main body tend to have $\Delta V >$ 60~$\mathrm{km\,s^{-1}}$, while smaller $\Delta V$
values are mostly seen for star clusters in the SMC main body. Moreover, the closer star 
clusters to the Sun, the larger their residual velocities, which could be a direct measure 
of the strength of the tidal interaction with the LMC 
\citep[mean heliocentric distance $\sim$49.9 kpc,][]{degrijetal2014}.

Figure~\ref{fig:velpos} shows the sky distribution of the studied star clusters with the
different outer SMC regions separated by dashed lines, namely: Northern Bridge (NB), 
Wing/Bridge (W/B), Southern Bridge (SB), West Halo (WH), and Counter Bridge (CB), 
respectively \citep{dias2016smc}.
Star clusters have been colored according to their dispersion velocities, 
those with larger $\Delta V$ values being mainly distributed in the outer SMC regions.
These regions are known to have been affected by LMC tides \citep[e.g.,][]{zivick2018proper,schmidt2020, dias2022viscacha, parisi2024,mackey2018substructures}, so that
the derived larger $\Delta V$ values could represent a measure of the strength of the
LMC tidal effects. For instance, L116, located in the Southern Bridge region, has a residual
 velocity of 225.73 $\mathrm{km\,s^{-1}}$ and is moving towards the LMC. In the 
West Halo, L4, 11, and 13  exhibit residual velocities
 greater than 110 $\mathrm{km\,s^{-1}}$, with velocity vectors oriented in the opposite 
direction to the LMC. Both the Wing/Bridge and the Northern Bridge have also star clusters with 
relative high residual velocities pointing towards the LMC (see Table~\ref{tab2}).
Star clusters located in the SMC main body or surrounding it generally have 
residual velocities $\Delta V$ $<$  60 $\mathrm{km\,s^{-1}}$. A 3D space view of the residual velocities 
is depicted in Figure~\ref{fig:cum3d}. As can be seen, the SMC is more elongated al ong the $x$ 
axis (approximately parallel to the SMC line-of-sight), with increasing residual 
velocities from its center out to its outskirts.

To characterize the kinematics of clusters in different substructures with a possible tidal origin, 
we analyze the dispersion of the 3D components of their residual velocities and compare them
to the total dispersion. Following the work of \citet{watkins2024mass}, we introduce the kinematic 
anisotropy in the SMC framework as follows:

\begin{equation}
A_i = \frac{\sigma^2(\Delta V_i)}{\sigma^2(\Delta V_x) + \sigma^2(\Delta V_y) + \sigma^2(\Delta V_z)}
\label{eq:anisotropy}
\end{equation}

\noindent for $i$ = $x, y, z$

Figure~\ref{fig:A} shows the values of $A_x$, $A_y$, and $A_z$ as a function of the
galactocentric distance, for each of the SMC disk models proposed in \citet{piatti2026}. At first glance,  star clusters pertaining to the outer
regions of the SMC tend to show a larger anisotropy along the $x$ and $z$ axes, which suggests an overall agitated kinematics approximately parallel to the SMC line-of-sight and perpendicular to its plane.

\subsection{West Halo}
The West Halo was proposed by \citet{dias2016smc} as a substructure distant from the 
SMC main body, and confirmed by proper motion studies 
\citep{niederhofer2018vmc,piatti2021kinematics}. Moreover, \citet{tatton2020vmc}
 suggested that the West Halo could be the tidal counterpart of the SMC Bridge
\citep[see also][]{zivick2018proper}.

\begin{figure}[H]
  \centering
  \includegraphics[width=\columnwidth]{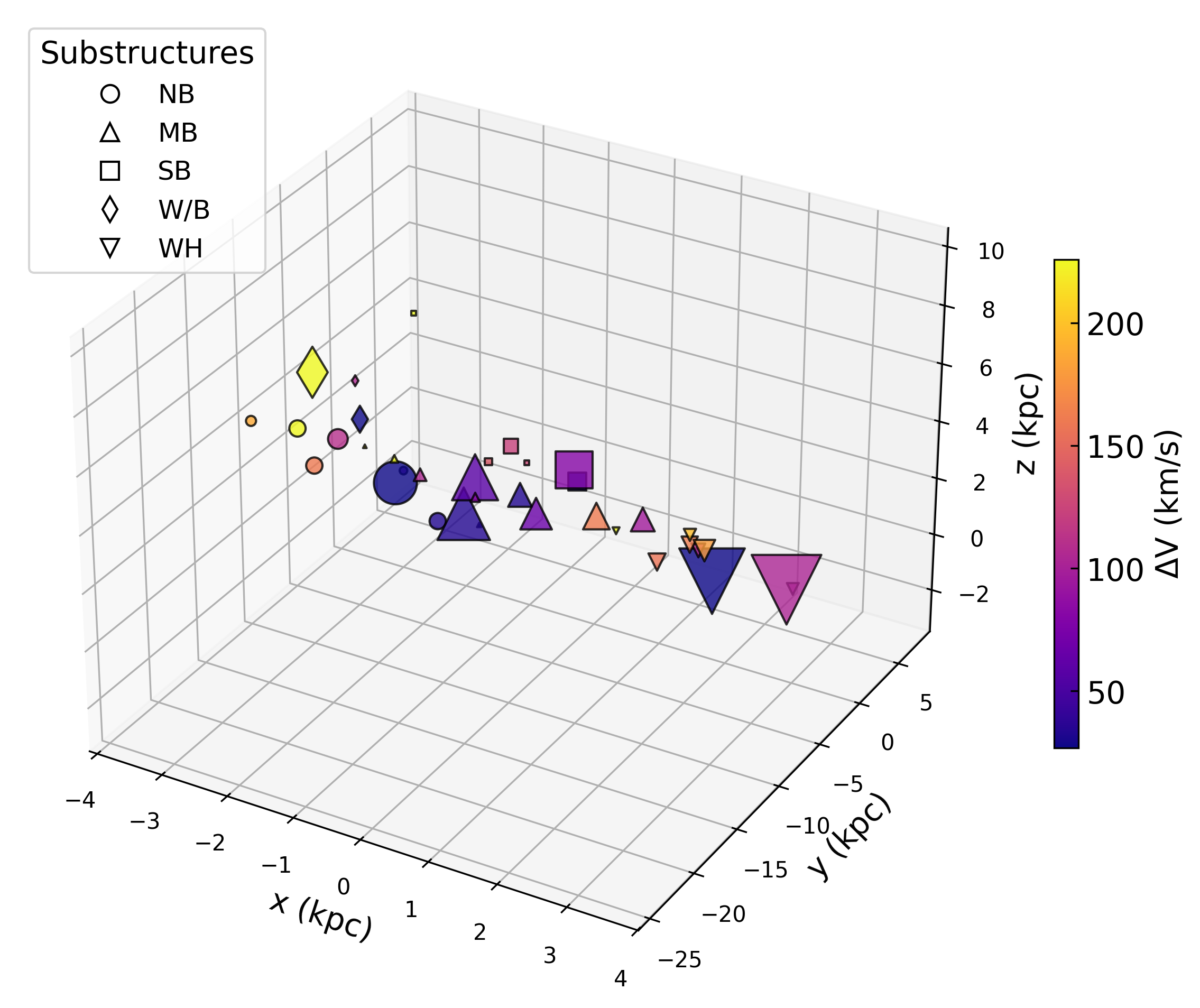}
  \caption{3D distribution of the studied star clusters. Star clusters projected onto
different SMC substructures are represented with different symbols, while their
colors correlate with their residual velocities. Symbol sizes are proportional to
 $N^{\mathrm{eff}}$.}
  \label{fig:cum3d}
\end{figure}

\begin{figure}
\includegraphics[width=\columnwidth]{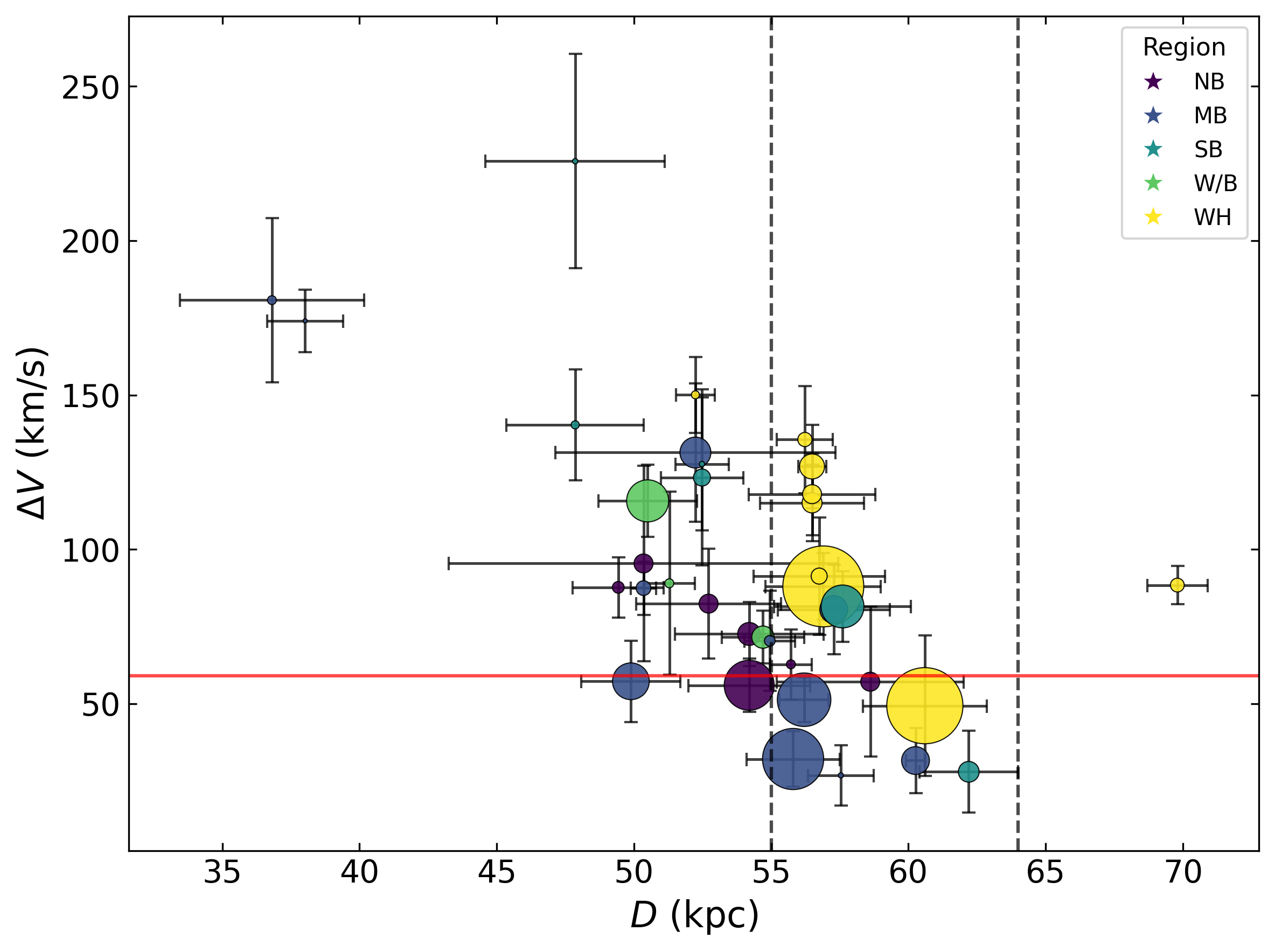}
\caption{Residual velocities as a function of the heliocentric distances of the studied 
star clusters. The vertical dashed lines represent the boundaries of the SMC main body \citep{piatti2021kinematics}, while the horizontal red line represents the lower residual velocity limit 
adopted in this work for star clusters located outside the SMC main body. 
Star clusters pertaining to different substructures \citep{dias2016smc}
are drawn with different colors as indicated in the top-right panel (NB=Northern Bridge, 
MB= Main Body, SB=Southern Bridge, W/B=Wing/Bridge, WH=West Halo). Symbol sizes are proportional to $N^{\mathrm{eff}}$.}
\label{fig:deltav}
\end{figure}

\begin{figure}
\includegraphics[width=\columnwidth]{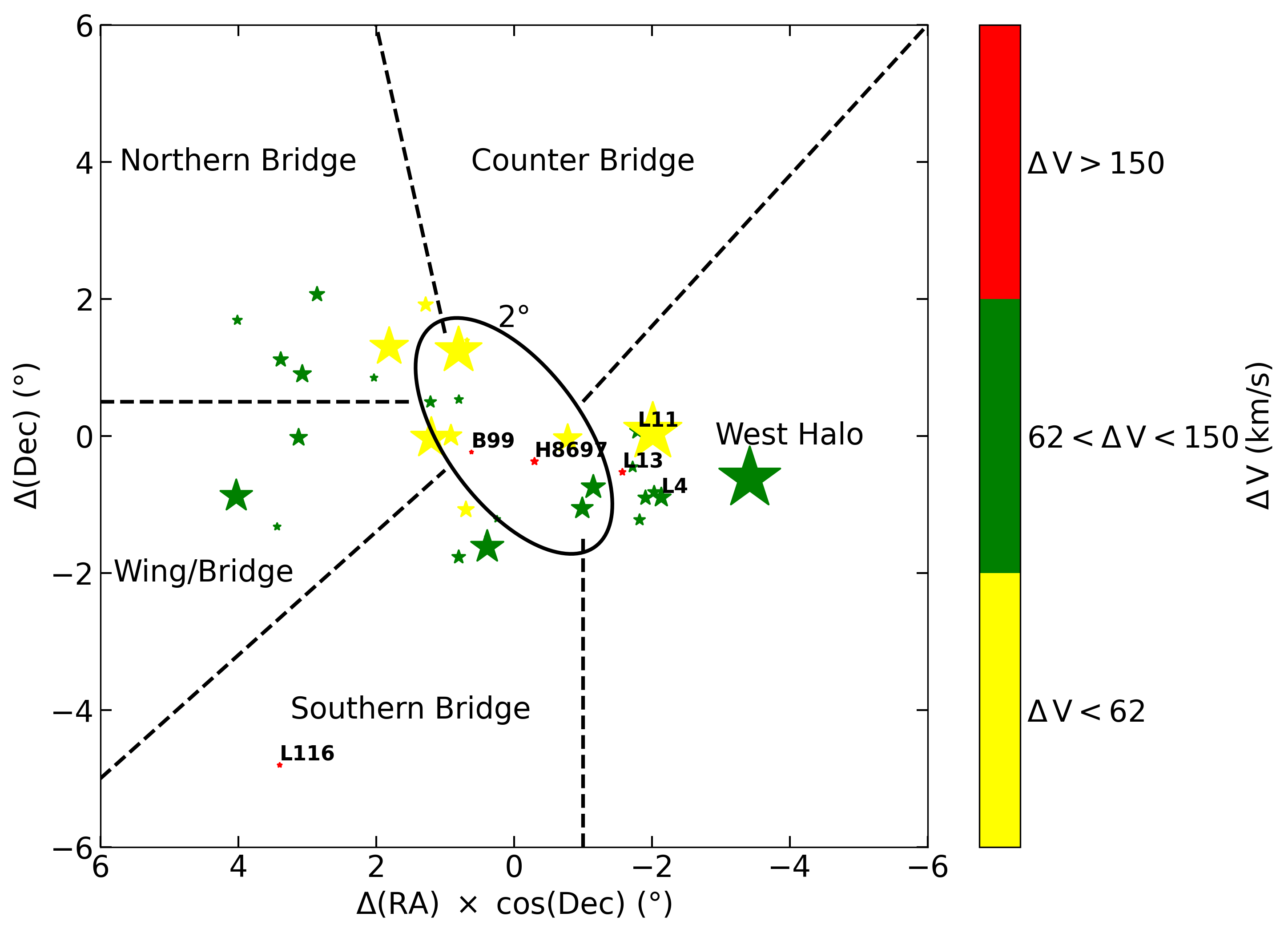}
\caption{Sky distribution of the studied star clusters, colored according
to their residual velocities. The dashed lines delimit the different outer SMC regions
\citep{illesca2025astrophysical}. Symbols sizes are proportional to $N^{\mathrm{eff}}$.}
\label{fig:velpos}
\end{figure}

\begin{figure*}[t]
\centering
\includegraphics[width=\textwidth]{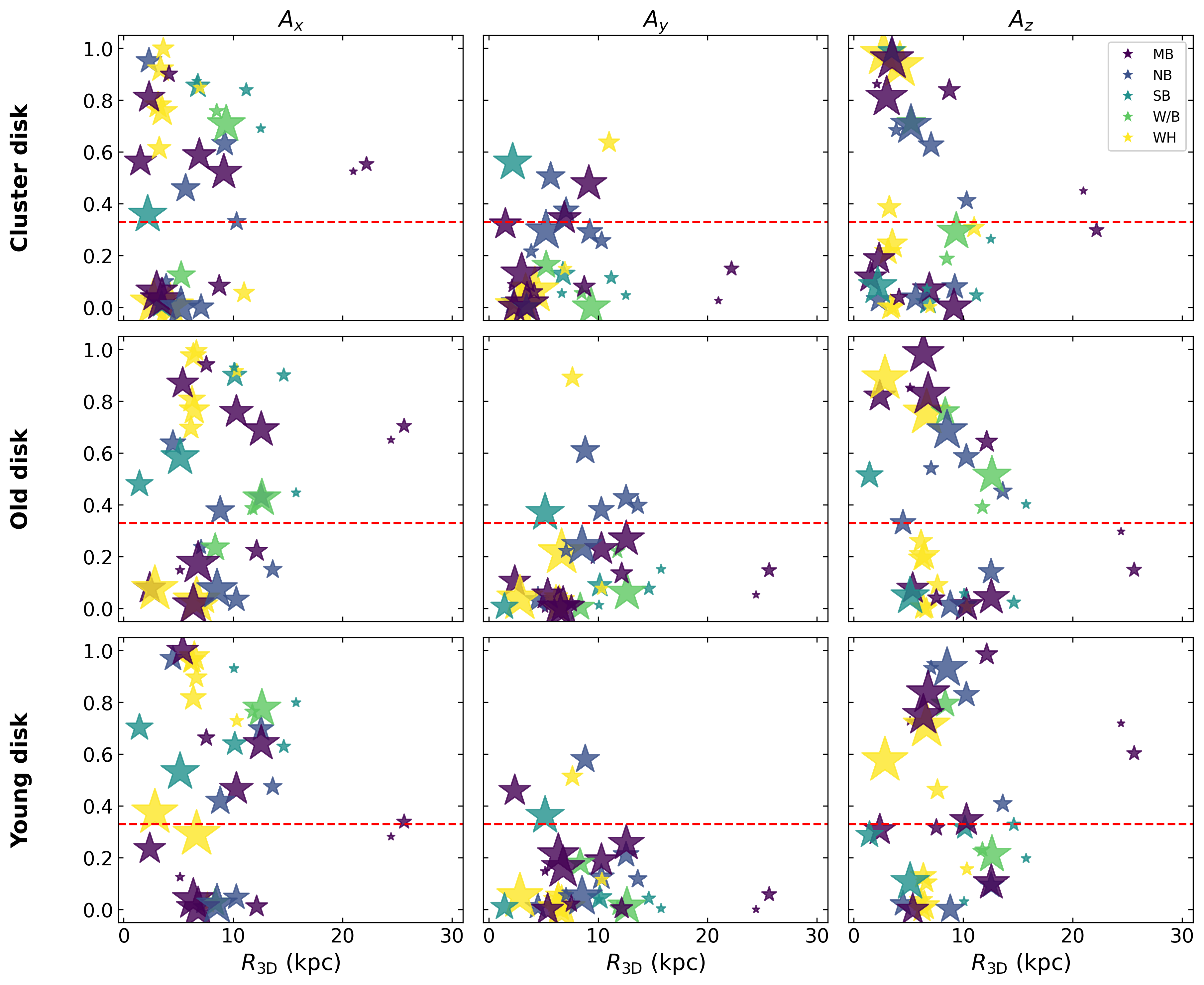}
\caption{Distribution of $A_x$, $A_y$, and $A_z$ as a function of distance from SMC center 
(${R_\mathrm{3D}}$).Anisotropy was estimated for each of the models studied in \citep{piatti2026}. As can be seen, the upper panel refers to the cluster disk model \citep{piatti2021kinematics}, the middle panel to the old disk model (age $> 2$ Gyr), and the lower panel to the young disk model (age $< 50$ Myr). Star clusters pertaining to different substructures \citep{dias2016smc}
are drawn with different colors as indicated in the top-right panel (NB=Northern Bridge, 
MB= Main Body, SB=Southern Bridge, W/B=Wing/Bridge, WH=West Halo). The red line ($A_i$ = 0.33, 
$i=x,y,z$) represents the expected value for an isotropic motion. Symbol size is proportional 
to $N^{\mathrm{eff}}$.}
\label{fig:A}
\end{figure*}

We obtained $A_x$ = 0.72, $A_y$ = 0.17, and  $A_z$ = 0.11, and a depth in the
spatial distribution of star clusters of $\sim $17 kpc, which point to a
clear elongation and predominant dispersion of motions along the $x$ axis 
(see Figure~\ref{fig:cum3d}). These outcomes reinforce the hypothesis that the West Halo 
is a dispersed  and disturbed substructure, possibly originated from a detachment of 
the SMC main body  \citep{dias2022viscacha}.

\subsection{Bridges and Wing}

For the Wing/Bridge region we obtained $A_x$ = 0.68, $A_y$ = 0.09, and $A_z$ = 0.23,
suggesting  that the star clusters are moving towards the LMC, as is also the case of
star clusters in the Southern Bridge ($A_x$ = 0.51, $A_y$ = 0.06, and $A_z$ = 0.44).
Four out of the six star clusters analyzed in this latter region have residual velocities larger
than the threshold value found in Figure~\ref{fig:deltav} (62~$\mathrm{km\,s^{-1}}$)
and heliocentric distances smaller than 53 kpc, which could be indicating escaping 
motions. On the other hand, star clusters in the Northern Bridge 
show a predominant motion dispersion perpendicular to the SMC plane 
($A_x$ = 0.16, $A_y$ = 0.29, and $A_z$ = 0.55). Three of them are located close to the boundary 
of the SMC main body, while the other four are placed at heliocentric distances smaller 
than \(51\,\mathrm{kpc}\). One again, the correlation between the amplitude of the residual 
velocities and the heliocentric distances reinforces their tidal origin \citep{piatti2022revisiting,sakowskaetal2024}.

\subsection{Main Body}

The studied star clusters projected on to the SMC main body span  $\sim$ 23.5 kpc of
heliocentric distance,  B99 and H86-97 being the closer star clusters
to the Sun ($D$ $<$ 39 kpc). These two star clusters have $\Delta V$ $>$ 170~$\mathrm{km\,s^{-1}}$,
which highlight from those physically occupying the SMC main body ($\Delta V$ $<$ 
60~$\mathrm{km\,s^{-1}}$).

\subsection{Kinematics under different SMC disk models}

As previously noted by \citet{piatti2026}, the estimation of $\Delta V$ depends on the adopted SMC rotation disk. Therefore, a comprehensive analysis of the kinematics of the studied star clusters requires considering different rotation disk models. \citet{dhanush2025unraveling} used {\it Gaia} DR3 data to derive kinematic parameters for different SMC star populations. From young to old star populations, they found a change in the SMC disk inclination from $\sim82^\circ$ to $\sim58^\circ$, and in the position angle of the line of nodes (LON) from $\sim180^\circ$ to $\sim240^\circ$.
Following the three SMC rotation disk models analyzed in \citet{piatti2026} (see Table~\ref{tab4}), we computed, for each kinematic scenario, the corresponding $\Delta V$ and the anisotropy along each SMC axis using a Monte Carlo approach. The relations between anisotropy and the distance of each cluster from the SMC center for the three disk models are shown in Fig.~5.

From the estimated global anisotropy, we obtain for the \textit{old disk} model $A_x = 0.61$, $A_y = 0.15$, and $A_z = 0.23$. For the model of \citet{piatti2021kinematics} we find $A_x = 0.63$, $A_y = 0.13$, and $A_z = 0.24$, while for the \textit{young disk} model we obtain $A_x = 0.72$, $A_y = 0.04$, and $A_z = 0.24$. 

These values indicate a more dispersed and dynamically perturbed kinematics along the line of sight (the $x$-axis) in the \textit{young disk} scenario. In contrast, the \textit{old disk} and \citet{piatti2021kinematics} models show a kinematic behavior with lower dispersion, although the dominant agitation still occurs along the line of sight of the galaxy.
The results under the \textit{young disk} scenario are fully consistent with the findings reported by \citet{piatti2026} and \citet{dhanush2025unraveling}, where star clusters exhibit a gradient in kinematic agitation as their distances from the SMC center increase. On the other hand, when analyzing the results obtained for the \citet{piatti2021kinematics} and \textit{old disk} models, we do not find a well-behaved kinematic distribution along the three axes of the galaxy, as might be expected considering that our cluster sample has a mean age of $\sim3$\,Gyr.

In this context, it is important to examine several key aspects. First, the clusters selected for this study are mostly located in external SMC substructures. Therefore, although the aforementioned models may capture the average agitation of older clusters, the presence of clusters in regions such as the West Halo or the Southern Bridge introduces a level of perturbation so high that their velocities exceed any average rotational behavior. For instance, the cluster L116 in the Southern Bridge exhibits a $\Delta V$ of 225.73 km\,s$^{-1}$, moving toward the LMC. These parameters likely place it outside any disk orbit, even a perturbed one.

Another important aspect is that, although the \citet{piatti2021kinematics} and \textit{old disk} models adopt different geometries compared to the young-cluster model, their geometries are still inferred from present-day observations. In other words, they do not fully represent the original disk geometry at the epoch when these clusters formed or when they were affected by the tidal forces of the LMC. Furthermore, we used individual heliocentric distances in the equations used to derive the velocities of each cluster. This provides additional robustness to the determination of $\Delta V$ and the corresponding anisotropy.

It is therefore likely that the parameters of the \citet{piatti2021kinematics} and \textit{old disk} models do not accurately reflect the magnitude of the kinematic agitation affecting old clusters located in the outer regions of the SMC. Our work does not aim to settle this debate, but rather to highlight the complexity involved in addressing the kinematics of the SMC.

\section{Conclusions}

The SMC is currently understood to be gravitationally bound to the LMC. Their interaction has left 
imprints on the SMC’s formation and evolution. Star clusters are fundamental building blocks of
 any galaxy, so it is reasonable to expect that they may contain valuable information about the
SMC  dynamical history.

In this work, we analyzed 36 star clusters in the SMC to derive their 3D velocities, with 
the aim of exploring 
the relationship between the star cluster kinematics and the tidal forces affecting
the SMC, particularlly in the SMC’s outer regions. We used proper motions from {\it Gaia} DR3, 
radial velocities taken from the literature, and our derived heliocentric distances. From these data, 
we derived 3D velocities and their residual velocities. Our main findings can be summarized as follows:

\begin{itemize}
\item The lower threshold for the residual velocities of star clusters located in
outer SMC regions is $\Delta V$ $\approx$ 60 $\mathrm{km\ s^{-1}}$, in very good agreement
with the value derived by \citet{piatti2021residual}. Star clusters belonging to the
SMC main body mostly show lower $\Delta V$ values, thus confirming a more tightly disk-like 
kinematics.

\item We performed an anisotropy analysis for different SMC disk models \citep{piatti2026}, based on recent findings by \citet{dhanush2025unraveling} linking the kinematics of the SMC with the age of the analyzed stellar sample. Although we found kinematic differences for each disk model, we also found certain regularities in relation to the kinematics and external substructures of the SMC: the West Halo, the Wing/Bridge,the Northern and the Southern Bridges show a preference for larger kinematic dispersion along 
the $x$ axis (approximately parallel to the SMC line-of-sight) and perpendicular to the disk, while star clusters in the SMC main body retains some amount of coherent rotation.

\item Our heliocentric distances \citep{piatti2023depth,illesca2025astrophysical} allowed us to construct a 
more realistic internal reference frame for the SMC. We thus report a line-of-sight depth 
for the studied star cluster sample of $\sim$ 25 kpc.

\item Building a 3D map of the SMC from the derived positions of each star cluster,
 combined with residual velocities and membership of star clusters to different SMC’s substructures, 
enabled us to identify spatial–velocity dispersion correlations.

\item The subregion-by-subregion analysis leads to an overall kinematic picture
of the SMC with kinematically hot outer regions, a pattern consistent with tidal models and 
recent close-encounter scenarios between both Magellanic Clouds \citep{rathore2024precise}.
\end{itemize}

\begin{acknowledgements}
We thank the referee for the thorough reading of the manuscript and timely suggestions to improve it.
This work has made use of data from the European Space Agency (ESA) mission Gaia 
(https://www.cosmos.esa.int/gaia), processed by the Gaia Data Processing and Analysis Consortium
 (DPAC, https://www.cosmos.esa.int/web/gaia/dpac/consortium). 

Data for reproducing the figures and analyses in this work will be available upon request to 
the first author.

\end{acknowledgements}

\bibliographystyle{aa}
\bibliography{biblio} 

\begin{appendix}
\onecolumn
\section{Collected and derived kinematic parameters of star clusters}

\begin{table}[h!]
    \caption{Proper motions and radial velocities of the studied star clusters.}
    \label{tab1}
    \begin{tabular}{c c c c c c c}\hline\hline
Star cluster & pmra & pmdec & $N$ & $D$ & RV & Ref.\\ 
& (mas\,yr$^{-1}$) & (mas\,yr$^{-1}$) & & (kpc) & (km\,s$^{-1}$) & \\\hline
B99 & $0.78\pm0.05$ & $-1.21\pm0.04$ & 1 & 38.02 & $159.20\pm2.60$ & 1\\
B168 & $0.94\pm0.09$ & $-1.15\pm0.09$ & 3 & 52.72 & $141.70\pm4.60$ & 2\\
BS121 & $0.82\pm0.06$ & $-1.23\pm0.04$ & 11 & 60.26 & $164.10\pm4.20$ & 3\\
BS188 & $1.25\pm0.08$ & $-1.35\pm0.07$ & 2 & 50.35 & $120.30\pm3.50$ & 2\\
H86-97 & $0.80\pm0.16$ & $-1.26\pm0.03$ & 2 & 36.81 & $120.90\pm2.80$ & 3\\
HW31 & $0.57\pm0.06$ & $-1.23\pm0.05$ & 2 & 47.86 & $125.50\pm3.40$ & 4\\
HW41 & $0.79\pm0.05$ & $-1.35\pm0.05$ & 1 & 57.54 & $143.60\pm1.60$ & 4\\
HW47 & $0.56\pm0.12$ & $-1.18\pm0.06$ & 5 & 52.48 & $122.90\pm2.40$ & 3\\
HW56 & $0.99\pm0.11$ & $-1.27\pm0.10$ & 2 & 58.61 & $157.70\pm5.40$ & 2\\
HW84 & $1.22\pm0.03$ & $-1.23\pm0.05$ & 3 & 49.43 & $135.60\pm1.50$ & 5\\
HW86 & $1.19\pm0.11$ & $-1.28\pm0.16$ & 2 & 51.29 & $143.80\pm1.60$ & 5\\
L1 & $0.58\pm0.01$ & $-1.53\pm0.01$ & 38 & 56.90 & $145.30\pm1.60$ & 6\\
L4 & $0.38\pm0.05$ & $-1.30\pm0.03$ & 9 & 56.49 & $140.20\pm1.60$ & 5\\
L6 & $0.50\pm0.05$ & $-1.33\pm0.03$ & 5 & 56.75 & $142.30\pm2.80$ & 5\\
L7 & $0.50\pm0.03$ & $-1.13\pm0.03$ & 6 & 56.49 & $131.40\pm2.60$ & 5\\
L8 & $0.67\pm0.13$ & $-1.32\pm0.04$ & 37 & 60.60 & $135.10\pm0.70$ & 6\\
L9 & $0.43\pm0.06$ & $-1.12\pm0.03$ & 4 & 56.23 & $157.40\pm2.10$ & 7\\
L11 & $0.42\pm0.03$ & $-1.28\pm0.04$ & 7 & 56.49 & $126.28\pm1.60$ & 8\\
L12 & $0.58\pm0.03$ & $-1.28\pm0.03$ & 4 & 69.80 & $208.00\pm1.30$ & 7\\
L13 & $0.46\pm0.05$ & $-1.14\pm0.02$ & 2 & 52.24 & $109.50\pm3.10$ & 3\\
L17 & $0.62\pm0.03$ & $-1.12\pm0.03$ & 12 & 52.24 & $106.00\pm1.60$ & 5\\
L19 & $0.54\pm0.04$ & $-1.29\pm0.03$ & 11 & 57.28 & $152.70\pm2.10$ & 5\\
L27 & $0.76\pm0.04$ & $-1.46\pm0.04$ & 14 & 49.89 & $175.00\pm2.60$ & 5\\
L58 & $0.47\pm0.08$ & $-1.32\pm0.07$ & 1 & 52.48 & $121.00\pm9.30$ & 6\\
L68 & $0.71\pm0.06$ & $-1.24\pm0.04$ & 8 & 62.20 & $143.70\pm0.83$ & 8\\
L100 & $0.81\pm0.05$ & $-1.17\pm0.05$ & 2 & 55.72 & $145.80\pm1.40$ & 2\\
L108 & $1.09\pm0.03$ & $-1.37\pm0.03$ & 7 & 54.20 & $95.00\pm4.00$ & 5\\
L110 & $0.79\pm0.02$ & $-1.18\pm0.02$ & 7 & 54.70 & $178.80\pm3.00$ & 5\\
L113 & $1.33\pm0.02$ & $-1.22\pm0.02$ & 17 & 50.50 & $171.80\pm4.50$ & 5\\
L116 & $1.63\pm0.09$ & $-1.10\pm0.07$ & 1 & 47.86 & $153.44\pm2.55$ & 8\\
NGC~339 & $0.65\pm0.03$ & $-1.21\pm0.03$ & 20 & 57.60 & $103.30\pm2.35$ & 8\\
NGC~361 & $0.83\pm0.03$ & $-1.28\pm0.02$ & 28 & 55.80 & $161.18\pm1.24$ & 8\\
NGC~416 & $0.90\pm0.03$ & $-1.19\pm0.04$ & 7 & 50.35 & $155.00\pm0.75$ & 9\\
NGC~419 & $0.87\pm0.04$ & $-1.22\pm0.02$ & 27 & 56.20 & $171.48\pm2.53$ & 8\\
NGC~458 & $0.89\pm0.01$ & $-1.23\pm0.02$ & 24 & 54.20 & $149.00\pm0.85$ & 9\\
OGLE~133 & $0.67\pm0.07$ & $-1.25\pm0.03$ & 3 & 54.95 & $145.40\pm3.20$ & 7\\\hline
\end{tabular}

\noindent Ref.:
(1) \citet{parisi2015ii}; 
(2) \citet{dias2021viscacha}; 
(3) \citet{dias2022viscacha}; \\
(4) \citet{de2022ii}; 
(5) \citet{parisi2009ii}; 
(6) \citet{piatti2021kinematics}; \\
(7) \citet{parisi2015ii}; 
(8) \citet{parisi2022ii}; 
(9) \citet{song2021dynamical}.
\end{table}

\begin{table*}[h!]
    \caption{Space velocity components of the star clusters.} 
\label{tab2}   
\begin{tabular}{cccccc}\hline\hline
Star cluster & $V_x$ & $V_y$ & $V_z$ & $V_\mathrm{rot}$ & $V_\mathrm{rot,3D}$ \\
& (km\,s$^{-1}$) & (km\,s$^{-1}$) & (km\,s$^{-1}$) & (km\,s$^{-1}$) & (km\,s$^{-1}$)\\
\hline
B99 & $-125.42 \pm 11.17$ & $-31.88 \pm 4.12$ & $-118.54 \pm 9.28$ & $129.48 \pm 11.16$ & $175.72 \pm 12.30$ \\
B168 & $-5.55 \pm 23.75$ & $-37.01 \pm 9.62$ & $-67.50 \pm 21.07$ & $43.78 \pm 11.85$ & $81.06 \pm 22.03$ \\
BS121 & $4.90 \pm 17.51$ & $5.22 \pm 5.73$ & $-26.49 \pm 11.25$ & $17.01 \pm 10.08$ & $33.22 \pm 10.74$ \\
BS188 & $59.35 \pm 48.49$ & $-52.00 \pm 12.72$ & $-41.72 \pm 35.67$ & $89.04 \pm 28.49$ & $105.40 \pm 25.36$ \\
H86-97 & $-136.05 \pm 32.90$ & $-62.39 \pm 6.75$ & $-99.12 \pm 17.51$ & $150.52 \pm 29.56$ & $181.43 \pm 27.31$ \\
HW31 & $-125.82 \pm 19.05$ & $-42.04 \pm 6.53$ & $-50.92 \pm 16.59$ & $132.87 \pm 18.69$ & $142.99 \pm 20.63$ \\
HW41 & $-7.90 \pm 14.67$ & $-6.37 \pm 5.04$ & $1.31 \pm 12.27$ & $16.20 \pm 9.01$ & $20.34 \pm 9.07$ \\
HW47 & $-109.96 \pm 31.90$ & $-40.74 \pm 6.79$ & $-43.91 \pm 18.44$ & $118.34 \pm 28.45$ & $128.06 \pm 26.12$ \\
HW56 & $40.49 \pm 33.42$ & $-4.82 \pm 11.78$ & $-31.37 \pm 26.66$ & $46.54 \pm 27.44$ & $62.86 \pm 25.72$ \\
HW84 & $43.23 \pm 11.23$ & $-47.49 \pm 4.55$ & $-70.53 \pm 11.36$ & $65.02 \pm 6.58$ & $96.45 \pm 8.55$ \\
HW86 & $58.75 \pm 28.59$ & $-33.70 \pm 13.15$ & $-64.93 \pm 37.13$ & $70.66 \pm 24.19$ & $100.82 \pm 31.74$ \\
L1 & $-39.17 \pm 14.59$ & $17.39 \pm 4.46$ & $75.78 \pm 10.67$ & $44.15 \pm 10.97$ & $88.88 \pm 5.07$ \\
L4 & $-127.63 \pm 13.73$ & $-1.58 \pm 3.58$ & $38.86 \pm 10.19$ & $127.70 \pm 13.71$ & $133.79 \pm 14.41$ \\
L6 & $-90.34 \pm 19.52$ & $-0.41 \pm 5.94$ & $31.45 \pm 15.44$ & $90.55 \pm 19.46$ & $97.46 \pm 17.53$ \\
L7 & $-120.02 \pm 13.70$ & $-28.50 \pm 4.68$ & $-16.75 \pm 11.30$ & $123.41 \pm 13.98$ & $124.96 \pm 14.82$ \\
L8 & $-22.55 \pm 36.77$ & $-8.21 \pm 6.83$ & $24.10 \pm 21.25$ & $37.46 \pm 23.89$ & $48.77 \pm 25.06$ \\
L9 & $-141.07 \pm 17.54$ & $-2.09 \pm 3.98$ & $-19.77 \pm 11.07$ & $141.14 \pm 17.51$ & $143.06 \pm 16.58$ \\
L11 & $-114.83 \pm 12.48$ & $-18.75 \pm 4.96$ & $28.90 \pm 12.73$ & $116.42 \pm 12.79$ & $120.82 \pm 10.81$ \\
L12 & $-5.71 \pm 11.97$ & $74.86 \pm 3.82$ & $40.08 \pm 10.40$ & $76.02 \pm 3.88$ & $86.29 \pm 7.94$ \\
L13 & $-144.55 \pm 13.09$ & $-53.64 \pm 3.73$ & $-21.08 \pm 7.52$ & $154.32 \pm 11.92$ & $155.98 \pm 11.30$ \\
L17 & $-106.79 \pm 27.42$ & $-63.28 \pm 7.31$ & $-44.07 \pm 19.80$ & $124.52 \pm 26.66$ & $132.63 \pm 30.96$ \\
L19 & $-82.68 \pm 14.56$ & $4.75 \pm 4.88$ & $8.38 \pm 13.14$ & $82.99 \pm 14.36$ & $84.57 \pm 13.60$ \\
L27 & $-51.56 \pm 14.45$ & $19.87 \pm 5.29$ & $-16.74 \pm 12.83$ & $56.04 \pm 12.20$ & $59.60 \pm 13.50$ \\
L58 & $-121.29 \pm 22.03$ & $-28.47 \pm 11.21$ & $1.15 \pm 19.87$ & $125.21 \pm 21.38$ & $126.81 \pm 21.21$ \\
L68 & $-14.89 \pm 19.37$ & $-7.21 \pm 4.99$ & $-2.26 \pm 14.21$ & $22.41 \pm 13.11$ & $26.70 \pm 12.97$ \\
L100 & $-20.69 \pm 12.90$ & $-23.89 \pm 4.88$ & $-51.16 \pm 12.05$ & $33.38 \pm 8.68$ & $61.51 \pm 12.96$ \\
L108 & $45.88 \pm 17.29$ & $-63.48 \pm 6.20$ & $-5.23 \pm 14.39$ & $80.04 \pm 8.16$ & $81.49 \pm 8.13$ \\
L110 & $-26.78 \pm 8.89$ & $6.66 \pm 4.08$ & $-67.68 \pm 8.52$ & $28.16 \pm 8.00$ & $73.58 \pm 9.85$ \\
L113 & $87.17 \pm 13.24$ & $-16.73 \pm 4.99$ & $-91.18 \pm 9.71$ & $89.01 \pm 12.55$ & $128.32 \pm 4.82$ \\
L116 & $169.15 \pm 41.16$ & $-57.11 \pm 7.06$ & $-157.03 \pm 27.13$ & $179.25 \pm 38.53$ & $241.56 \pm 25.70$ \\
NGC~339 & $-59.48 \pm 14.72$ & $-52.02 \pm 4.88$ & $-10.85 \pm 13.48$ & $79.44 \pm 13.16$ & $81.13 \pm 14.21$ \\
NGC~361 & $-11.44 \pm 11.35$ & $-0.18 \pm 3.56$ & $-31.16 \pm 8.67$ & $14.04 \pm 8.67$ & $34.89 \pm 10.09$ \\
NGC~416 & $-31.07 \pm 8.23$ & $-23.59 \pm 3.46$ & $-77.48 \pm 8.94$ & $39.41 \pm 6.98$ & $87.12 \pm 9.75$ \\
NGC~419 & $-4.12 \pm 10.27$ & $3.62 \pm 3.57$ & $-52.34 \pm 7.18$ & $10.54 \pm 6.10$ & $53.75 \pm 7.05$ \\
NGC~458 & $-8.68 \pm 11.80$ & $-20.57 \pm 3.60$ & $-49.36 \pm 9.41$ & $24.55 \pm 6.93$ & $55.23 \pm 11.19$ \\
OGLE~133 & $-58.82 \pm 17.77$ & $-14.26 \pm 4.68$ & $-25.96 \pm 9.26$ & $61.01 \pm 16.70$ & $67.43 \pm 14.64$ \\
\hline
    \end{tabular}
\end{table*}

\begin{table*}[h!]
    \caption{Residual velocity components of star cluster.} 
    \label{tab3}   
    \begin{tabular}{c c c c c c c c}\hline\hline
Star cluster & $\Delta V_x$ & $\Delta V_y$ & $\Delta V_z$ & $\Delta V$ & $R_{\mathrm{plane}}$ & $R_{\mathrm{3D}}$ & Projected\\
& (km\,s$^{-1}$) & (km\,s$^{-1}$) & (km\,s$^{-1}$) & (km\,s$^{-1}$) & (kpc) & (kpc) & region$^a$\\\hline
B99 & 126.96 & 28.99 & 115.20 & $174.01\pm10.17$ & 22.85 & 24.42 & MB \\
B168 & 15.23 & 42.29 & 62.64 & $82.42\pm17.82$ & 9.84 & 10.19 & NB \\
BS121 & 1.03 & 9.20 & 23.43 & $31.59\pm10.58$ & 2.13 & 2.38 & MB \\
BS188 & 49.76 & 48.39 & 37.79 & $95.44\pm31.67$ & 11.62 & 12.44 & NB \\
H86-97 & 134.72 & 64.34 & 97.35 & $180.74\pm26.56$ & 24.03 & 25.63 & MB \\
HW31 & 125.34 & 43.46 & 40.36 & $140.34\pm17.94$ & 13.35 & 14.61 & SB \\
HW41 & 11.68 & 13.11 & 9.69 & $26.80\pm9.70$ & 5.09 & 5.14 & MB \\
HW47 & 110.25 & 39.83 & 29.78 & $123.27\pm28.54$ & 8.74 & 10.11 & SB \\
HW56 & 34.88 & 13.06 & 22.94 & $57.13\pm24.18$ & 4.48 & 4.48 & NB \\
HW84 & 32.24 & 45.49 & 66.44 & $87.70\pm9.72$ & 12.67 & 13.47 & NB \\
HW86 & 54.62 & 23.93 & 52.51 & $89.02\pm29.65$ & 9.90 & 11.63 & B \\
L1 & 28.27 & 18.87 & 79.87 & $88.01\pm10.73$ & 6.33 & 6.55 & WH \\
L4 & 119.26 & 4.32 & 42.21 & $126.86\pm13.48$ & 5.97 & 6.39 & WH \\
L6 & 82.13 & 2.69 & 34.65 & $91.35\pm18.97$ & 5.72 & 6.10 & WH \\
L7 & 112.23 & 31.82 & 13.42 & $117.89\pm13.28$ & 5.88 & 6.31 & WH \\
L8 & 15.40 & 2.45 & 28.75 & $49.36\pm22.73$ & 2.81 & 2.81 & WH \\
L9 & 133.92 & 7.99 & 15.36 & $135.59\pm17.39$ & 6.01 & 6.59 & WH \\
L11 & 108.53 & 13.38 & 33.21 & $115.06\pm12.43$ & 6.02 & 6.20 & WH \\
L12 & 2.64 & 75.63 & 43.16 & $88.43\pm6.18$ & 7.22 & 7.68 & WH \\
L13 & 138.07 & 54.11 & 18.71 & $150.06\pm12.31$ & 9.69 & 10.31 & WH \\
L17 & 101.65 & 66.98 & 41.19 & $131.37\pm22.46$ & 9.55 & 10.28 & MB \\
L19 & 78.05 & 1.48 & 13.55 & $80.54\pm14.49$ & 4.79 & 5.36 & MB \\
L27 & 48.36 & 22.18 & 14.88 & $57.27\pm13.16$ & 11.87 & 12.57 & MB \\
L58 & 120.95 & 29.49 & 13.71 & $127.65\pm21.50$ & 8.88 & 10.05 & SB \\
L68 & 15.73 & 4.80 & 9.96 & $27.95\pm13.25$ & 0.39 & 1.43 & SB \\
L100 & 28.79 & 23.59 & 48.21 & $62.77\pm11.37$ & 6.65 & 7.03 & NB \\
L108 & 36.52 & 59.31 & 1.20 & $72.61\pm10.35$ & 8.14 & 8.74 & NB \\
L110 & 33.63 & 15.73 & 60.58 & $71.57\pm8.53$ & 7.33 & 8.28 & W/B \\
L113 & 82.00 & 6.08 & 80.17 & $115.71\pm11.72$ & 10.79 & 12.47 & W/B \\
L116 & 168.16 & 53.20 & 137.76 & $225.73\pm34.71$ & 11.94 & 15.40 & SB \\
NGC~339 & 59.01 & 53.42 & 2.81 & $81.51\pm11.40$ & 3.98 & 5.12 & SB \\
NGC~361 & 15.57 & 6.61 & 24.47 & $32.09\pm8.92$ & 6.63 & 6.79 & MB \\
NGC~416 & 36.30 & 23.62 & 75.58 & $87.45\pm8.67$ & 11.42 & 12.14 & MB \\
NGC~419 & 8.42 & 8.55 & 48.48 & $51.30\pm7.17$ & 5.80 & 6.35 & MB \\
NGC~458 & 16.12 & 25.08 & 45.45 & $55.98\pm8.64$ & 8.16 & 8.49 & NB \\
OGLE~133 & 63.00 & 16.71 & 23.85 & $70.39\pm16.24$ & 7.16 & 7.54 & MB \\
\hline
    \end{tabular}

\noindent $^a$ NB = Northern Bridge, W/B = Wing/bridge, SB = Southern Bridge,
MB = Main Body, and WH = West Halo \citep{dias2016smc}
\end{table*}

\begin{table*}[h!]
  \centering
  \caption{SMC rotation disk models.}
  \label{tab4}
  \begin{tabular}{lccc}
    \toprule
    \textbf{Parameter} &
    \textbf{Cluster disk} &
    \textbf{Young disk (age $<$ 50 Myr)} &
    \textbf{Old disk (age $>$ 2 Gyr)} \\
    \midrule
    SMC center RA ($^\circ$) &
      $13.30\pm0.10$ & $13.05$ & $13.05$ \\
    SMC center Dec ($^\circ$) &
      $-72.85\pm0.10$ & $-72.83$ & $-72.83$ \\
    SMC center distance (kpc) &
      $59.0\pm1.5$ & $62.44\pm0.47$ & $62.44\pm0.47$ \\
    SMC center $\mathrm{pmra}$ (mas\,yr$^{-1}$) &
      $0.75\pm0.10$ & $-0.743\pm0.027$ & $-0.743\pm0.027$ \\
    SMC center $\mathrm{pmdec}$ (mas\,yr$^{-1}$) &
      $-1.26\pm0.05$ & $-1.233\pm0.012$ & $-1.233\pm0.012$ \\
    SMC center systemic velocity (km\,s$^{-1}$) &
      $150.0\pm2.0$ & $145.6\pm0.1$ & $145.6\pm0.1$ \\
    SMC disk inclination ($^\circ$) &
      $70.0\pm10.0$ & $81.9\pm0.7$ & $58.4\pm1.4$ \\
    SMC disk position angle LON ($^\circ$) &
      $200.0\pm30.0$ & $185.7\pm3.7$ & $207.6\pm2.3$ \\
    SMC disk rotation velocity (km\,s$^{-1}$) &
      $25.0\pm5.0$ & $10.0\pm5.0$ & $10.0\pm5.0$ \\
    \bottomrule
  \end{tabular}
\end{table*}
\end{appendix}

\end{document}